# Tunneling anomalous Hall effect in the nanogranular CoFe-B-Al-O films near the metal-insulator transition

V.V. Rylkov[1,2 (a)], S.N. Nikolaev[1], K.Yu. Chernoglazov[1], V.A. Demin[1], A.V. Sitnikov[3], M.Yu. Presnyakov[1], A.L. Vasiliev[1], N.S. Perov[4], A.S. Vedeneev[5], Yu.E. Kalinin[3], V.V. Tugushev[1], A.B. Granovsky[2,4]

[1]National Research Centre "Kurchatov Institute", 123182 Moscow, Russia
[2]Institute of Applied and Theoretical Electrodynamics RAS, 127412 Moscow, Russia
[3]Voronezh State Technical University, 394026 Voronezh, Russia
[4]Faculty of Physics, Lomonosov Moscow State University, 119991 Moscow, Russia
[5]Kotel'nikov Institute of Radio Engineering and Electronics RAS, 141190 Fryazino, Moscow Region, Russia

___________________________________

E-mail: (a) vvrylkov@mail.ru

## Abstract

We present results of experimental studies of structural, magneto-transport and magnetic properties of CoFe-B-Al-O films deposited onto a glass ceramic substrate by the ion-beam sputtering of the target composed of $Co_{40}Fe_{40}B_{20}$ and $Al_2O_3$ plates. The system consists on the strained crystalline CoFe metallic nanogranules with the size 2-5 nm which are embedded into the B-Al-O oxide insulating matrix. Our investigations are focused on the anomalous Hall effect (AHE) resistivity $\rho_{AHE}$ and longitudinal resistivity $\rho$ at $T$=5-200 K on the metallic side of metal-insulator transition in samples with the metal content $x$=49-56 at.%, that nominally corresponds to $(Co_{40}Fe_{40}B_{20})_x(Al_2O_3)_{100-x}$ in the formula approximation. The conductivity at $T > 15$ K follows the $\ln T$ behavior that matches a strong tunnel coupling between nanogranules. It is shown that the scaling power-laws between $\rho_{AHE}$ and $\rho$ strongly differ, if temperature $T$ or metal content $x$ are variable parameters: $\rho_{AHE}(T) \propto \rho(T)^{0.4-0.5}$ obtained from the temperature variation of $\rho$ and $\rho_{AHE}$ at fixed $x$, while $\rho_{AHE}(x)/x \propto \rho(x)^{0.24}$, obtained from measurements at the fixed low temperature region (10-40 K) for samples with different $x$. We qualitatively describe our experimental data in the frame of phenomenological model of two sources of AHE e.m.f. arising from metallic nanogranules and insulating tunneling regions, respectively, at that the tunneling AHE (TAHE) source is strongly shunted due to generation of local circular Hall currents. We consider our experimental results as the first experimental proof of the TAHE manifestation.

**I. Introduction**

In spite the anomalous Hall effect (AHE) was first explained in 1954 [1] and intensively studied during last years [2] several questions on the relative importance and specific features of basic mechanisms of AHE in different systems are still under strong debates. Nowadays the renewed interest to AHE is related with its common origin with direct and inverse spin Hall effects [3,4], which are key phenomena in spintronics, orbitronics and magnonics.

The AHE is the most clearly pronounced in magnetic materials (ferromagnetic metals and semiconductors, granular metal–insulator nanocomposites, etc.) with strong spin-orbit interaction (SOI) [2]. Their Hall resistivity $\rho_H$ is described by the sum of two terms

$$\rho_H = R_0 B + 4\pi R_s M, \tag{1}$$

where the first term describes normal Hall effect (NHE) induced by the Lorentz force and the second term characterizes AHE related to SOI, $M$ is the magnetization component perpendicular to the film plane, $B$ is the magnetic induction component in this direction, $R_0$ and $R_s$ are so called NHE and AHE coefficients, respectively.

One of the most interesting lines in the AHE research in magnetic systems is the study of a relation between anomalous component of the Hall resistivity $\rho_{AHE} = 4\pi R_s M$ and longitudinal resistivity $\rho_{xx} = \rho$, i.e. the so-called scaling behavior $\rho_{AHE} \propto \rho^n$, where $n$ is the power law index determined by one of the other physical mechanism of AHE [2]. The scaling relation in this form (or its equivalent for conductivities: $\sigma_{AHE} = \rho_{AHE} / \rho^2 = \rho_{AHE} \sigma^2 \propto \sigma^\gamma$ with $\gamma \approx 2-n$) is widely used in the literature, if impurity concentration or temperature are variable parameters [2]. The simple scaling behavior is well established for homogeneous magnetic systems with the one type of impurity. For example, in low-resistivity magnetic metals with a not very strong impurity scattering $n=1$ in the case of skew scattering mechanism, while $n=2$ in the case of side-jump or intrinsic mechanism. With increasing of impurity scattering potential in high-resistivity (so called "dirty") magnetic metals the index $n$ decreases to $n \approx 0.4$ [2]. Nevertheless, there are numerous cases when scaling relation is not maintained and for some heterogeneous systems considerable deviations from the scaling law were reported (see, e.g., [5-8] and references therein).

Rather frequently, interpretation of the AHE data is contradictory and intricate in complex magnetic materials, and the most of investigations of scaling relation between $\rho_{AHE}$ and $\rho$ were performed for systems which do not exhibit metal-insulator transition (MIT) [2, 5-8]. However, some complex structures, in particular granular nanocomposites in which it is possible to vary their resistivity by several orders of magnitude (from good metal to insulator) by changing the metal volume fraction are the most convenient systems for investigation of the scaling relation (if it exists) and other features of the AHE behavior in different metallic and insulating regimes.

To explain experimental results in such the structures, Efetov et al. [9] considered a theoretical model of dense-packed ferromagnetic granules coupled to each other by tunneling contacts in metallic regime and found that there is no scaling relation between transverse and longitudinal resistivity. In this theory, the AHE regime arises only inside the granules. However, Vedyaev et al. [10] showed that AHE may arise inside tunneling barriers due to influence of SOI on the scattering of electrons on the intergranular located impurities or a Rashba spin-orbit coupling within the tunneling barrier layer. Recently, other TAHE mechanism was considered in Ref. [11], caused by the interfacial SOI and resulting in a "skew" electron tunneling even in the absence of impurities.

Earlier in [12], the AHE theoretical model was considered for the hopping transport systems: the authors obtained the scaling law with $n=0.5$ if the impurity concentration is a variable parameter. The basic finding of this theory was that AHE originates from the influence of SOI on correlated hopping between triads of impurities in the percolation network. That is similar to the NHE in the hopping regime which appears in triads of impurities under the influence of magnetic field on the interference between the amplitude for a direct and indirect (second-order) transition [13]. Detailed calculations of AHE for a hopping between triads of impurities under SOI in the percolation network was carried out in Ref.[14] and it was obtained the scaling with $n$ between 0.67 and 0.24 depending on the specific features of hopping transport. However, the above theory of AHE in a percolation network is valid, strictly speaking, only for a hopping regime in diluted systems with isolated magnetic impurities, but not for nanocomposites with dense-packed ferromagnetic granules.

An interesting feature of granular nanocomposites consists in unusual behavior of their conductivity near MIT. It was predicted in [15, 16] that in the close vicinity of MIT on the metallic side when the tunneling conductance between granules $G_t$ is much larger than the quantum conductance $G_q = 2e^2/\hbar$ ($g = G_t/G_q >> 1$) the conductivity should follow the ln$T$ behaviour ($\sigma \propto$ ln$T$). Just under these conditions the AHE resistivity $\rho_{AHE}$ does not depend on longitudinal resistivity, i.e. $n \approx 0$ [5]. Recently, such unconventional scaling law with $n \approx 0$ when conductivity follows logarithmic law $\sigma \propto$ ln$T$ has been demonstrated for Ni-SiO$_2$ nanocomposites by varying Ni content [17] (in spite of $n \approx$ 0.6-0.7 far from MIT [17, 18]). The correlation between $\rho_{AHE}$ and $\rho$ when the temperature is a variable parameter has been not studied in [17].

Parametric dependences of $\rho_{AHE}(\rho)$ vs $T$ and $x$ have been previously studied in Fe-SiO$_2$ nanocomposites on the dielectric side of MIT where the dependence $\ln\sigma \propto (T_0/T)^{1/2}$ comes true [19, 20]. It was proved that temperature variation of the $\rho_{AHE}(\rho)$ dependence follows the power law $\rho_{AHE} \propto \rho(T)^n$, $n = 0.44$-$0.59$ [20]. Meanwhile, at the Fe content variation the $\rho_{AHE}(\rho)$ function revealed strongly nonmonotonic behavior: the Hall effect conductance shows a flattening in the MIT vicinity and a tendency to reach a new plateau at $T_0$ increasing [20].

The universal scaling factor with $n\sim 0.5$ was observed for $p$-type insulating Ga$_{1-x}$Mn$_x$As ($x\sim 0.014$) for films with different hole concentration [21]. On the other hand, in Ga$_{1-x}$Mn$_x$As on the metallic side of MIT ($x \geq 0.05$) the power-laws $\rho_{AHE}(T) \propto [\rho(T)]^2$ and $\rho_{AHE}(x) \propto [\rho(x)]^{0.5}$ were established at the temperature and Mn content variations, respectively [22]. This difference in the power-law index was explained by a specific dependence of magnetization of metallic cluster near MIT on its conductivity; $M \propto \sigma_{xx}^{1.5}$ at low temperatures significantly smaller than the Curie temperature $T_C$.

Below we present our results on structural, magneto-transport and magnetic properties of nanogranular CoFe-B-Al-O thin films with an excess oxygen vacancies in oxide matrix focusing on scaling relations between Hall resistivity and longitudinal resistivity at $T$=5-200 K. Note that nanostructures based on oxides like AlO$_z$ ($z < 1.5$) possess a resistive switching phenomena caused by existence of oxygen vacancies; these phenomena can be used for memristor implementation, synapse simulation and creation of new type smart devices [23, 24].

Previously it was shown that similar nanocomposites do not exhibit well-defined percolation threshold [25] and therefore are suitable objects to study AHE in strongly disordered metallic regime. We consider metallic regime with the ferromagnetic alloy content $x$ =49-56 at.% close to MIT when conductivity follows the $\ln T$ behavior and shows that the scaling power-law differs if temperature $T$ or content $x$ are variable parameters: $n = 0.4$-$0.5$, obtained from the temperature variation of $\rho$ and $\rho_{AHE}$ measured for each sample and $n \approx 0.24$, obtained from measurements at fixed low temperature (10-40 K) for samples with different content. We attribute our data to a complex nature of AHE as a result of action of two parallel e.m.f. sources caused by SOI inside the metallic granules and insulating tunneling barrier between granules.

**II. Samples**

The granular films under study were produced using the ion-beam sputtering of the composite targets onto glass-ceramic substrates at growth temperature not exceeding 100 °C. The targets include the parent metallic alloy $Co_{40}Fe_{40}B_{20}$ and twelve aluminum oxide ($Al_2O_3$) plates placed onto the surface of this metal. The special target design makes it possible to obtain composite systems with the relative content of the metallic and insulating phases continuously varying in a wide range $x$ = 25-60 at.% in a single technological cycle [26]. The thickness of the produced samples was about $d$ = 2.7 μm. The elemental composition of the films was determined by energy dispersive X-ray spectroscopy using an Oxford INCA Energy 250 unit attached to a JEOL JCM-6380 LV scanning electron microscope.

Above we adduce the metal fraction of grown nanocomposite, approximating it structure by the formula $(Co_{40}Fe_{40}B_{20})_x(Al_2O_3)_{100-x}$. We will further use this approximation for the composition characterization because it allows definitely find the $x$ value by data of energy dispersive X-ray spectroscopy as well as gets information about chemical content of the composition target. On the other hand, the data of structural studies (see below) shows that considerable part of the B atoms in nanocomposite is appeared to be outside the CoFe granules. The enthalpy of the BO oxide formation (+0.04 eV/molecule) is much less as compared to the enthalpy of the AlO oxide formation (+0.95 eV/molecule) but the binding energy of BO molecule (8.4 eV) is much large than of the AlO molecule (5.0 eV) [27]. For this reason the boron atoms outside CoFe granules are energetically more favorable to form the BO oxide while residual oxygen to form $AlO_z$ ($z < 1.5$) oxide. In the limit when all boron atoms are outside the metal granules the nanocomposite of $(CoFe)_x(BO)_y(AlO_z)_{100-x-y}$ type could be formed. The content of metal phase herewith decreases. For instance, if in the case of $(Co_{40}Fe_{40}B_{20})_x(Al_2O_3)_{100-x}$ nanocomposite the $x$ value is ≈ 50 at.%, then the nanocomposite transformation to $(CoFe)_x(BO)_y(AlO_z)_{100-x-y}$ leads to the $x$ ≈ 40 at.% and $z$ ≈ 1.

After producing composites, we used photolithography for preparing the samples having the standard double-cross shape to measure the electrical conductivity and Hall effect resistance. The conduction channel had the width $w$=1.2 mm, the length $l$=4 mm with the distance between potential probes $l_p$=1.4 mm.

## III. Structural characterization by electron microscopy

### A. Experimental details

The cross-section samples for transmission electron microscopy (TEM) and scanning transmission electron microscopy (STEM) were prepared by a $Ga^+$ focused ion beam (FIB) in a scanning electron-ion microscope HeliosNanoLab ™ 600i (FEI, USA) equipped with Pt and W gas injection systems (GIS) and with a micromanipulator Omniprobe 200 (Omniprobe, USA). To protect the sample surface during specimen preparation, the protective Pt layer with the thickness of ~1.5 µm was deposited by the $e^-$- beam. Standard FIB procedure was used for specimen preparation: the 2 µm lamella was cut by focused $Ga^+$ beam with the energy of $E = 30$ keV and a current of $I = 6.5$ nA, than with the help of the micromanipulator the lamella was removed and attached to Cu semicircle Omniprobe (Omniprobe, USA) by W deposition. Final thinning of the specimen to electron transparency was made by $Ga^+$ beam with the energy of $E=2$ keV, and current of $I=28$ nA. The assessed specimen thickness was less than 5 nm. The specimens were studied in a Titan 80-300 TEM/STEM (FEI, USA) with a spherical aberration ($C_s$ probe) corrector at an accelerating voltage of 300 kV. The microscope was equipped with a field emission cathode (Schottky), SuperTwin objective lens with spherical aberration coefficient of 1.2 mm, energy dispersive X-ray (EDX) spectrometer (EDAX, USA) and a high angle annular dark-field electron detector (HAADF) (Fischione, USA). The EDX microanalysis including elemental mapping was additionally performed in a Tecnai Osiris TEM/STEM (FEI, USA) with attached Super-X EDX system (Bruker, USA) at an accelerating voltage of 200 keV. For the image processing Digital Micrograph (Gatan, USA) software and TIA (FEI, USA) were used.

### B. Results

The HAADF STEM images of samples $(Co_{40}Fe_{40}B_{20})_x(Al_2O_3)_{1-x}$ with $x$=46 and 57 at.% were presented in Fig.1a and b, respectively. We failed to find noticeable difference in particle sizes (lie between $a \approx 2$-5 nm) in these two samples. The selected area electron diffraction (SAED) pattern from one of the samples is presented in Fig.2a and intensity histogram along the white line is shown in Fig.2b. It demonstrates the peaks correspondent to three interplanar spacings: 2.02, 1.25 and 0.8 Å. These spacings match to the distances close to $d(110)$, $d(211)$ and $d(222)$ in base centered cubic (bcc) FeCo alloy with unit cell constant $a_c = 0.28486$ nm (Space Group Im-3m) [28]. However (200) bcc reflection in our case is absent.

Note that similar missing of (200) bcc reflection was observed previously in Fe-Cr-N alloy after mechanical milling [29] as well as in the study of milled Fe-Cr-X compounds, where X was N, C and B [30]. After HRTEM investigations the authors proposed that bow-shaped deformation of the crystal planes of the particles with the size of few nm occurred along the (110) bcc slip planes. Such deformation may cause distortion of the bcc (200) planes and drastic decrease of (200) reflection. That proposition was confirmed by simulation of diffraction pattern from the distorted $Fe_{50}Co_{50}$ bcc model consisting of 1000 unit cells [30].

We checked the number of different compounds, including oxides and borides but all these compounds exhibited the crystal structure with less symmetry, thus the diffraction patterns demonstrate comparatively large number of peaks. Thus we rule out all these compounds from our consideration.

Close inspection of the SADP, presented in Fig. 2a indicate the presence of amorphous halo and that could be linked with the presence of amorphous oxide and boride substance between FeCo particles.

Since two samples demonstrated similar microstructure, below we will consider in details only specimen with $x$=57 at.%. As it was shown in Fig.1b, the HAADF STEM image demonstrated the areas of bright and dark contrast and it can be proposed that the areas with bright contrast correspond to FeCo particles as with high Z number. The areas with the dark contrast could be related to Al and B oxides. To prove that suggestion we performed the EDX mapping and the results are presented in Fig.3(a-e). There is unambiguous match of Fe and Co distribution and these areas correspond to the bright areas in HAADF STEM image. In contrary, the darker areas correspond to more intense signal from Al and O and theses areas are pointed arrows in Fig. 3(a-e). Boron is the lightest element which can be detected by EDX and the efficiency of B registration is relatively low. So, the B elemental map was not informative and we performed the EDX line scan across bright and dark areas (see the bottom of Fig.4, the scan line is marked by the arrow). Again, the correspondence between Fe and Co from one hand and Al and O from the other is very clear. The B distribution was not conclusive and it should be studied in more details with other spectroscopic methods. We only could speculate that B was distributed more or less uniformly both in Fe-Co and Al-O regions with slight excess at the boundaries between them (Fig.4).

The bright field (BF) HR STEM image of the sample is shown in Fig.5a. The lattice image of particles was clearly visible. They are more pronounced in the dark areas, which corresponded to FeCo particles. These lattice fringes correspond to the $(110)_{FeCo}$ crystal planes. Surprisingly, these lattice fringes clearly demonstrated the presence of the texture in relatively large areas, more than 20 nm and that was confirmed by Fast Fourier Transforms from the HR STEM images and one of these is shown in Fig. 5b. The angle between two <110> maxima was less than $90^0$, but we attributed that to lattice image distortions during scanning because of charging effects. On small areas we did observe image of the crystal planes intersecting under the $90^0$ (see enlarge image in Fig.5c). The BF HR STEM image of lattice, obtained from the sample with $x$=46 at.% is even more impressive. However, the areas having bright contrast, which correspond to Al (and B) oxides looks more amorphous (not shown here).

## IV. Transport and magnetic properties
### A. Longitudinal conductivity

The dependence of the resistivity $\rho(x)$ on the content $x$ of $(Co_{40}Fe_{40}B_{20})_x(Al_2O_3)_{100-x}$ samples in the MIT vicinity ($x$ = 46.5-59.2 at.%) measured at the temperature $T$=77 K is shown in Fig. 6a. The temperature dependence of the longitudinal conductivity $\sigma(T)$ for samples with different $x$ is shown on Fig. 6b. In linear scale, the resistivity (Fig. 6a) starts to increase gradually at $x \leq 56$ at.%, while at $x \leq 49$ at.% the $\rho(x)$ dependence starts to be exponentially strong. In the range $x$=(49-56) at.% and $T$ > (10-15) K conductivity is well described by the law $\sigma(T) \propto \ln T$ (Fig.6b) that corresponds to the metallic regime with strong tunnel coupling between granules [15, 16]. Note that description of the $\sigma(T)$ dependence by the logarithmic law outside this range (at $x$ > 56 at.% and $x$ < 49 at.%) is noticeably worse (Fig.6b). Furthermore, at $x \leq x_c \approx 47$ at.% ($\rho_c$ = (2-3)·$10^{-2}$ Ω·cm) this law changes to $\ln\sigma \propto (T_0/T)^{1/2}$ related to hopping conductivity (see insert of Fig.6b) [16]. It means that MIT in our case occurs near $x_c \approx 47$ at.%; on the other hand, classical percolation threshold takes place at $x_p \approx 56$ at.%. The probable reason of such a noticeable difference between $x_c$ and $x_p$ seems to be connected with the percolation transition spreading which is due to a nonvanishing conductivity of dielectric component of composite [31]. In our case the conductivity of the bad oxidized B-Al-O matrix can be significant owing to low barriers between granules. For example, under bad oxidized conditions the barrier height can be less than 1 eV in tunnel junctions type of metal/$AlO_x$/metal (Me/$AlO_x$/Me) [32, 33]. On the other hand for optimally oxidized Al-O matrix the barrier reaches ≈3 eV for Me/$AlO_x$/Me tunnel junctions [34, 35]. Below we focus our attention on the region of compositions with $x$ = (49-56) at.% and resistivity $\rho$ of $10^{-2}$ - 2·$10^{-3}$ Ω·cm at 77 K, that corresponds to the case of high resistivity dirty metals [2].

**B. Hall effect and magnetization**

In heterogeneous systems such as a percolation medium, noticeable voltage always presents between the Hall probes even in the absence of magnetic field. It appears due to asymmetry of the percolation network (see [36] and reference therein) and can be presented as a product of the "asymmetry resistance" $R_a$ and longitudinal current $I_x$ through the sample. To separate the components $R_H$ and $R_a$ of the Hall resistance the measurements were carried out for two opposite directions of magnetic field $B$; their values were calculated as $R_H(B) = [R_{xy}(B)-R_{xy}(-B)]/2$ and $R_a(B) = [R_{xy}(B)+R_{xy}(-B)]/2$, where $R_{xy} = V_y/I_x$ is the transverse resistance. The $R_a$ value varies not only due to magnetoresistive effect, but also due to percolation cluster reconstruction in the magnetic field [36]. In our case the "asymmetry resistance" $R_a$ is much larger than the Hall resistance: at the saturation of magnetization typical values of $R_H$ for samples studied at low temperatures are in the range of 0.01-0.03 Ω (Fig.7a). Meanwhile, the $R_a$ value could sometimes exceed 0.1 Ω. For example, in the sample with $x$ = 49 at. % the value of $R_a$ reaches 0.45 Ω at $T$ = 10 K (Fig.7b).

The additional hard problem in measurements is connected to a spin glass-like behavior of the samples caused probably by the partial oxidation of the granule surface and formation of a thin film of the CoO-type antiferromagnetic oxide [37]. The glass-like behavior of nanocomposites under investigation is evident, in particular, from the long-time resistivity relaxation after the magnetic field switching (see insert of Fig.7b).

To suppress negative influence of the large "asymmetry resistance" and uncertainness caused by the spin glass-like behavior on the results of the AHE temperature dependence investigations we used the following sequence of measurements: (i) the sample was cooled down to 10 K, magnetic field +0.5 T was applied and the sample was heated up to 200K; (ii) magnetic field was increased up to +1.5 T and the sample was kept 5 min in this field and then we started measurements of $R_{xy}(T,+1.5T)$ during 40 min of slow cooling down to 10K; (iii) we applied magnetic field -0.5 T in opposite direction and performed the same procedures (i) and (ii) to measure $R_{xy}(T,-1.5T)$; iv) then we measured the field dependence of $R_{xy}(B)$ at 10 K and compared the saturation value $R_H(10K)$ with that obtained by using $R_{xy}(T,+1.5T)$ and $R_{xy}(T,-1.5T)$. If these two values of $R_H(10K)$ coincide well with each other the procedure for determination of $R_H(T)$ is considered as correct and to analyze the temperature dependencies we use the values of $R_H(T)$ in saturated state ($B$=1.5 T).

Usually, both the AHE conductivity $\sigma_{AHE}$ and resistivity $\rho_{AHE}$ are linear functions of the magnetization $M(x,T)$ that need to be considered at the scaling relation studies. The $M(x,T)$ measurements were performed with SQUID magnetometer (Quantum Design PPMS-9T) at 5-300 K in magnetic fields up to 7 T. The temperature dependence of $M(x,T)$ measured at 1.5 T is strong at 5-20 K and is not saturated in this field (Fig.8a) probably because of a large amount of paramagnetic Co and Fe atoms in the oxide B-Al-O matrix and/or superparamagnetic granules belonging to the dead and ragged ends of the percolation cluster [38]. Note that the data of structural measurements also indicate the presence of Co and Fe atoms in this matrix (Fig. 4). On the other hand, at $T \geq 25$ K the magnetization $M(x,T)$ is practically independent of $T$ in comparison with the resistance $\rho(T)$ (Fig.8b). Due to this reason, one could neglect the $M(x,T)$ temperature variation and assume that the $\rho_{AHE}(T)$ dependence is mainly determined by the $\rho(T)$ dependence. Also notice that the $M(x,T)$ variation with the metal content in range of $x = (49-56)$ at.% should not be large. Below, analyzing the scaling dependence $\rho_{AHE}$ vs $\rho(x)$, we will fit the $M(x,T)$ dependence on the metal content in this range by a linear function of $x$.

The dependence of the AHE conductivity $\sigma_{AHE}$ on the longitudinal conductivity $\sigma_{xx}$ at $B = 1.5$ T for samples with $x=49$ and 53 at.% at different temperatures is shown in Fig.9. These data support the universal scaling law, since the obtained index $\gamma =1.55-1.61$ is very close to the widely accepted value $\gamma =1.6$ (or $n=0.4$) for dirty metals with $\rho \geq 10^{-4}$ $\Omega$·cm [2]. On the other hand, the power-law index $n$ in the $\rho_{AHE}(x)/x \propto \rho(x)^n$ relation changes to the surprisingly low value $n=0.24$ if the metal content $x$ is a variable parameter (Fig. 10). Note that in this case the $n$ value is independent of the temperature at $T < 40$ K (Fig. 10); that testifies a reasonableness of the approximation $M(x) \propto x$.

## V. Discussion
### A. Crucial remarks on the longitudinal conductivity

Above presented conductivity results for CoFe-B-Al-O nanocomposites based on strongly nonstoichiometric oxide show a relatively broad percolation threshold spreading region near the MIT vicinity, $\delta x_p = (x_p - x_c) \sim 7$ at.% (Fig. 6a). In this region, experimental temperature dependence of the conductance $\sigma$ is well fitted by the logarithmic function: $\sigma = \sigma_0 + \beta \cdot \ln T$, where $\sigma_0$ and $\beta$ are the fitting parameters. This dependence is typical for a granular metallic system at large tunnel conductance between granules and not very low temperature [15, 16], when a weak localization effects are suppressed. Its physical origin is not connected with the system dimension, but is only due to renormalization of the Coulomb interaction by impurity scattering processes, that affects the quasiparticle tunneling between granules [15, 16].

Estimation of the percolation threshold region for Ni-SiO$_2$ nanocomposites based on stoichiometric oxide gives, however, the much less value of $\delta x_p = (x_p - x_c) \sim 1$ at.% [17]. From this dissimilarity we can suppose a significant role of the oxide matrix in the conductivity of our system caused by the existence of oxygen vacancies (bad matrix oxidation) leading the low energy (less than 1 eV) tunneling barriers between metallic grains to appear. That is typical situation for the Me/AlO$_x$/Me tunnel junctions containing bad oxidized AlO$_z$ regions ("hot spots") with an effective tunneling area of about 0.1-1 µm$^2$ [32, 33] which strongly exceeds in our case the granule cross-section (percolation network).

Another peculiarity of our CoFe-B-Al-O nanocomposites is a very small variation of the slope $\beta$ in the logarithmic temperature dependence of the conductance $\sigma$ when the metal content $x$ is varied in the percolation threshold spreading region, $x$=(49-56) at.% (see Fig. 6b). In contrast to the Ni-SiO$_2$ system, where $\beta$ has a change of ~ 2 times at $\delta x_p \sim 1$ at.% [17], in our case $\beta$ = 26-32 (Ω·cm)$^{-1}$ at $\delta x_p \sim 7$ at.%. According to the model [15, 16], below the percolation threshold when $g = G_t/(2e^2/\hbar) > 1$, where $g$ and $G_t$ are, respectively, normalized and non-normalized tunneling conductance, the conductivity of granular nanocomposite follows the law:

$$\sigma(T) = \sigma_0 \left(1 - \frac{1}{2\pi Dg} \ln\left[\frac{gE_c}{k_B T}\right]\right) = \sigma_0 \left(1 - \frac{1}{\pi kg} \ln\left[\frac{gE_c}{k_B T}\right]\right). \quad (2)$$

Here $D$ is the system dimension, $k$ is the coordinate number ($k$ = 6 for closely packed cubic lattice), $E_c$ is the Coulomb energy. The formula (2) is correct at $g\delta << T << E_c$, where $\delta$ is the mean energy level spacing in a single granule.

The connection between the fitting parameter $\beta$ and the model parameters in (2) is evident: $\beta = \sigma_0 \left(\frac{1}{\pi kg}\right)$. Assumption that parameter $\sigma_0$ is connected with the tunneling conductance $G_t$ by the relation $G_t \approx \sigma_0 \frac{L^2}{L} = \sigma_0 L$ [38] leads to the following expression (also see [16]):

$$\sigma_0 \approx \frac{2e^2 g}{\hbar L}, \quad (3)$$

where $L$ is the size of the percolation cluster cell. Substituting of the relation (3) into (2) gives:

$$\beta = \frac{2e^2}{hkL}. \quad (4)$$

In the case of two-component system of the metal-insulator type, the $L$ value is weekly dependent on the metal content $x$ in the percolation threshold spreading region [31]. According to the data of Fig.6 the $x$ decreasing actually means in our system an effective increasing of the intergranular distance $b$ and also a decreasing of the tunneling conductance $G_t$ at about constant percolation cluster cell $L$. Substituting the experimentally obtained slope value $\beta = 26\text{-}32$ $(\Omega\cdot\text{cm})^{-1}$ into (4) leads to the value of $L \approx 8$ nm that is of the order of the granule size (2-5 nm) estimated from the electron microscopy results. Note that at the approximation of closely packed granules when the granule size $a \gg b$, the size of the percolation cluster cell $L \approx a$ [16]. Thus we can conclude that conducting chains which form the percolation cluster in our nanocomposites at $x=$(49-56) at.% contain (1-2) tunnel junctions with intergranular distance ~1-2 nm.

**B. Qualitative model of the AHE behavior**

Conditions for appearance of metallic conductance $g = G_t/(2e^2/\hbar) > 1$ could be fulfilled even at relatively large distances between metallic granules, $b$ ~1-2 nm, if intergranular barriers formed by insulating B-Al-O oxide matrix are sufficiently low ($\leq 1$ eV). In this situation, the $V_{Hd}$ contribution to the Hall e.m.f. may arise in the intergranular regions, for example due to the mechanism [6] associated with electron tunneling between granules and SOI scattering of electrons on the paramagnetic Fe and Co atoms dissipated inside the matrix (Figs. 4 and 8a). At the same time, the $V_{Hg}$ contribution to the Hall e.m.f. caused by the granules themselves [5] also exists in the system. To qualitatively describe AHE in this complex structure, let us consider a simple phenomenological model with two Hall e.m.f. sources connected in parallel one with other (Fig 11a). Similar situation takes place in a macroscopic rectangular semiconductor sample [39], where the circular Hall current occurs in the vicinity of metallic electrodes (Fig.11b). That leads to a strong reduction of the potential drop $V_H$ measured between probes as compared to the Hall e.m.f. arising in the interelectrode regions, $V_{Hd} = R_{Hd} \cdot I_x$. The physical reason of this reduction is a shunting of the Hall e.m.f. by local circular Hall currents near the electrode surface. The geometry of such a system is characterized by geometrical parameter $b/a$: according to Ref.[39],

$$V_H/V_{Hd} \approx 0.65 \text{ at } b/a = 1 \text{ and } V_H/V_{Hd} \approx 0.3 \text{ at } b/a = 0.5, \qquad (5)$$

where $R_{Hd}$ is the Hall resistance of interelectrode (in our case intergranular) regions, $a$ is the size of metallic granule which plays a role of metallic electrode in our case.

Measured value of the Hall resistance $R_H = V_H/I_x$ can be estimated at $b/a \leq 1$ on the base of an equivalent circuit model taking into account combined influence of two Hall e.m.f. sources connected in parallel (see Fig.11c). Fig.11c illustrates effective Hall e.m.f. generator for a periodic net of tunnel junctions which contains: (i) the source of Hall e.m.f. in metallic granules, $V_{Hg} = R_{Hg} \cdot I_x$, with internal resistance $r_{gint}$, and (ii) the source of Hall e.m.f. in dielectric intergranular regions, $V_{Hd} = R_{Hd} \cdot I_x$, with internal resistance $r_{dint} \gg r_{gint}$. These sources are connected in parallel to each other through external resistance $r_{dext} \gg r_{gint}$. The effective Hall e.m.f. generator (see Fig.11c) generates the Hall e.m.f. $V_{Heff} = V_H = 1/2 \cdot [(\varphi_{g1} - \varphi_{g0}) + (\varphi_{d1} - \varphi_{d0})]$; both resistances $r_{dint}$ and $r_{dext}$ are obviously determined by the charge carriers transport in the intergranular region along it long side $a$ and in the transverse direction, respectively, i.e. $r_{dint} \propto a/(a \cdot b)$ and $r_{dext} \propto b/(a \cdot a)$. In the case of macroscopic samples when $V_{Hg} = 0$ it could be shown that both conditions (5) are fulfilled with an accuracy not worse than 10% at $r_{dint}/r_{dext} \approx 1/2\,(a/b)^2$ and $b/a \leq 1$. On this base we obtain the value of measured Hall resistance: $R_H(x,T) \approx R_{Hd}(x,T) \cdot (b/a)^2 + R_{Hg}(T)$. Obviously, in frame of this phenomenological approach we cannot specify the $R_H(x,T)$ behavior as a function of $x$ and $T$. For this reason, below in our estimations we will rely on the principal idea of the microscopic model of TAHE exposed in Ref. [10], while others microscopic models may be possibly used.

Generally speaking, there exist several factors of influence of the metal content $x$ on $R_H(x,T)$. The problem is that not only the total number of metal atoms, but their redistribution between granules and intergranular regions varies with $x$, even if we assume that $R_{Hg}(x,T)$ is independent of $x$. On the one hand, the value of $x$ may affect an effective energy barrier $U(x)$ between metallic granule due to redistribution of metal atoms between different electron and magnetic configurations inside and outside the granule. For simplicity, let us completely disregard this dependence and take into consideration only universal "geometrical" factor $[b(x)/a]^2$ in the Hall resistance dependence $R_H(x,T)$ on $x$ since the intergranular distance $b(x)$ evidently increases as $x$ decreases. Thus, in our approximation both sources, the intergranular tunneling AHE source and granular AHE source, contribute to the total AHE following the formula:

$$R_H(x,T) \sim R_{Hd}(x,T) \cdot [(x_p-x)/x_p]^2 + R_{Hg}(T). \qquad (6)$$

The factor $[(x_p-x)/x_p]^2$ reducing the component $R_{Hd}(x,T)$ in Eq.(6) reflects an aforementioned shunting of the TAHE source in our structure. Note that Eq. (6) signifies that the main reason of the AHE increase with the metal content decrease near MIT is not connected with *electron scattering* but is rather caused by a *percolation cluster topology*. Following the model described by Eq.(6), this topology manifests itself in an increasing of the TAHE contribution into $R_H(x,T)$ at the metal content decreasing near MIT. In fact, it is a main reason for different scaling power-laws between $\rho_{AHE}$ and $\rho$, if temperature $T$ or metal content $x$ are variable parameters.

Let us first consider the temperature dependence of $R_H(x,T)$ at fixed metal content $x$. Since the granules size is small (2-5 nm) one can expect very strong scattering by their interfaces and therefore the temperature behavior of the corresponding contribution $R_{Hg}(T)$ to $R_H(x,T)$ is the same as for dirty metals. In spite the theory of TAHE at finite temperatures is absent we can only speculate about the character of temperature dependence of the tunneling contribution $R_{Hd}(x,T)$ to $R_H(x,T)$. Large amount of paramagnetic impurities inside tunnel barriers, possible ferromagnetic order between their magnetic moments, the presence of superparamagnetic granules and granules size distribution, spin-flip processes, – all these factors makes the temperature dependence of $R_{Hd}(x,T)$ similar to that for dirty metals, or at least not stronger. So, our system at varied $T$ and fixed $x$ should behave similarly to an effective continuous dirty metal. Indeed, from Fig. 9 we can attribute to this system a scaling law with the index $\gamma$ =1.55-1.61 which is close to the widely accepted value $\gamma$ =1.6 (or $n$=0.4) for a dirty metal with $\rho \geq 10^{-4}$ $\Omega \cdot$cm [2]. Thus we conclude that our rough description is not so bad while formally it is not correctly justified.

At the same time, with a varied metal content dependence of $R_H(x,T)$ specified by above-stated shunting effect becomes more pronounced. Our measurements at the fixed low temperature $T$ and varied metal content $x$ show the different scaling relation between $\rho_{AHE}(x)$ and $\rho(x)$: $\rho_{AHE}(x)/x \propto \rho(x)^{0.24}$ (see Fig.10), that formally corresponds to $\gamma$ =1,76. This fact is not surprising since completely different mechanisms are responsible for concentration dependences of AHE and resistivity: a shunting effect does not play any role in resistivity.

To additionally verify our model we assume that below the percolation threshold the AHE variation with $x$ is completely determined by the TAHE contribution while at $x > x_p \approx$ 56 at.% the $R_H(x,T) \approx R_{Hg}(T)$. From Eq. (6) thus follows an estimation of the TAHE component, $R_{TAHE} \approx \Delta R_H =$ $[R_H(x,T) - R_{Hg}(T)] \approx R_{Hd}(x,T) \cdot [(x_p-x)/x_p)]^2 \propto R_{sd} \cdot M \cdot (\Delta x/x_p)^2$, where $R_{sd}$ is the TAHE coefficient, $\Delta x = x_p - x << x_p$. So, below the percolation threshold, an increment of the AHE resistivity normalized on the magnetization should follow the law $\Delta(\rho_{AHE}/x) \propto (\Delta x)^\beta$ with $\beta \approx 2$.

The $\Delta(\rho_{AHE}/x)$ vs $\Delta x$ dependence obtained for the low temperature region, $T \leq 25$ K, is shown in Fig. 12. It is clear that this dependence is well described by the power law with $\beta \approx 2$. We treat this result as a clear manifestation of the TAHE contribution in the total AHE. At first glance, however, the dependence in Fig. 12 is surprising: it indicates that the TAHE coefficient $R_{sd}$ is independent of the intergranular distance $b(x)$. We may speculate, however, that for our system with low energy barriers $U(x)$ between metallic granules the height and shape of $U(x)$ vary with $x$ due to redistribution of Fe and Co atoms between granules and intergranular regions. As a result, the decreasing and smoothing of $U(x)$ may partially compensate the effect of increasing of $b(x)$ on $R_{sd}$ at the metal content $x$ decreasing under $\Delta x << x_p$.

**Conclusion**

We considered metallic regime of the CoFe-B-Al-O nanocomposites with the metal content $x$ = 47-59 at.% in its formula approximation $(Co_{40}Fe_{40}B_{20})_x(Al_2O_3)_{100-x}$. We showed that the relation between the AHE resistivity $\rho_{AHE}(T,x)$ and longitudinal resistivity $\rho(T,x)$ does not follow the universal scaling law, if $T$ or $x$ are variable parameters: $\rho_{AHE}(T) \propto \rho(T)^{0.4 \div 0.5}$ ($\gamma = 1.5$-$1.6$), obtained from the temperature variation of $\rho$ and $\rho_{AHE}$ measured for each sample at fixed $x$ and $\rho_{AHE}(x)/x \propto \rho(x)^{0.24}$ ($\gamma = 1.76$), obtained from measurements at a fixed low temperature (10-40 K) for samples with different $x$. We explain our data by a sufficiently small ($\leq 1$ eV) height of insulating intergranular barriers appearing between metallic granules and the electron-impurity SOI scattering in these barriers leading to TAHE mechanism accompanying ordinary AHE mechanism provided by metallic granules. A logarithmic temperature dependence of longitudinal resistivity is observed for a relatively broad region of the metal content variation, and the manifestations of TAHE regime are revealed.

Finally, we give preference to the TAHE mechanism caused by electron SOI scattering on the 3d impurities inside tunneling barriers [10], since we observed strong paramagnetic signal (Fig. 8a) in our system. However, we also cannot *a-priori* exclude a possibility of the TAHE manifestation caused by the skew tunneling process [11] owing to a crystal lattice ordering in the relatively large scale ~20 nm for the CoFe-B-Al-O nanocomposites under study (Fig. 5).

**Acknowledgements:**

This work was financially supported by Russian Science Foundation, grant No. 16-19-10233. The work was partially done on the equipment of the Resource center of Kurchatov Institute Complex of NBICS-technologies. The authors acknowledge partial support from M.V. Lomonosov Moscow State university Program of Development for magnetic measurements.

**Figure captions**

Fig.1. HAADF STEM images of $(Co_{40}Fe_{40}B_{20})_x(Al_2O_3)_{1-x}$ nanocomposites with (a) $x$=46 and (b) $x$=57 at.%. Light areas correspond to Co-Fe metal clusters.

Fig.2. (a) The selected area electron diffraction (SAED) pattern from sample with $x$=46 at.%. (b) Intensity histogram along the white line shown in Fig. 2 (a).

Fig.3. (a) - HAADF STEM image of the sample with $x$=57 at.% and the elemental maps of (b) - Fe, (c) - Co, (d) – Al, (e) –O. Note the overlapping of Co and Fe elemental distribution in the brighter areas of the image (a) and Al with O in darker areas, which are pointed by arrows.

Fig.4. Fe, Co, Al, O and B elemental distribution along the arrow shown in HAADF STEM image in the bottom part.

Fig.5. Bright field high resolution STEM image of the sample with $x$=57 at.%. (b) The FFT from the image. (c) The enlarge image of the sample demonstrating lattice image with orthogonal {110} planes.

Fig.6. (a) Resistivity of $(Co_{40}Fe_{40}B_{20})_x(Al_2O_3)_{100-x}$ samples vs ferromagnetic alloy content near MIT ($x$ = 46.5-59.2 at.%) at $T$ = 77 K. (b) The temperature dependences of conductivity for samples with different ferromagnetic alloy content $x$ = 47-59 at.%.

Fig.7. (a) Magneto-field dependences of the Hall resistance for samples with $x$ = 47, 49 and 59 at.% at low temperatures $T$ < 25 K. (b) Temperature dependence of "asymmetry" resistance $R_a(T)$ for sample with $x$ = 49 at.%. Insert shows the relaxation curve of longitudinal resistance for this sample after application of 1 T field during ~ 1 min.

Fig. 8. (a) Temperature dependences of normalized magnetization $M$ for $(Co_{40}Fe_{40}B_{20})_x(Al_2O_3)_{100-x}$ films with $x$=49 and 56 at.%. (b) Normalized $M(T)$ dependence in comparison with normalized resistivity $\rho(T)$ dependence for sample with $x$=49 at.%. Insert shows the normalized $M(T)$ and $\rho(T)$ dependences in enlarged temperature scale ($T \approx$ 30-190 K).

Fig.9. Logarithmic plots of AHE conductivity $\sigma_{AHE}$ vs $\sigma_{xx}$ for samples with $x$=49 and 53 at.%.

Fig.10. Logarithmic dependence of normalized AHE resistivity ($\rho_{AHE}/x$) vs longitudinal resistivity $\rho$, obtained from measurements at a fixed low temperature in the range of 10-36 K for samples with different metal content $x$.

Fig.11. (a) Granular system with SOI scattering on defects in oxide matrix at electron tunneling between grains. Two connected in parallel sources of AHE e.m.f. are shown: the first source is caused by spin-orbit interaction in granules ($V_{Hg} = R_{Hg} \cdot I_x$) and the second occurs inside the tunneling barrier regions ($V_{Hd} = R_{Hd} \cdot I_x$); the current $I_x$ flows through the neighboring granules, $R_{Hg}$ and $R_{Hd}$ are the Hall resistances of granules and dielectric interlayer between them. (b) Schematic draw of tunneling junction between granules illustrating an emergence of circular currents at formation of the Hall effect in dielectric interlayer. (c) Equivalent scheme of a periodic network of tunnel junctions (resistances) with the two local Hall e.m.f. generators; $r_{dint} \gg r_{dext} \gg r_{gint}$, $V_{Heff} = \frac{1}{2}[(\varphi_{g1} - \varphi_{g0}) + (\varphi_{d1} - \varphi_{d0})]$.

Fig.12. Variation of normalized AHE resistivity $\Delta(\rho_{AHE}/x)$ vs $\Delta x = (x_p - x)$ at $T \leq 25$ K.

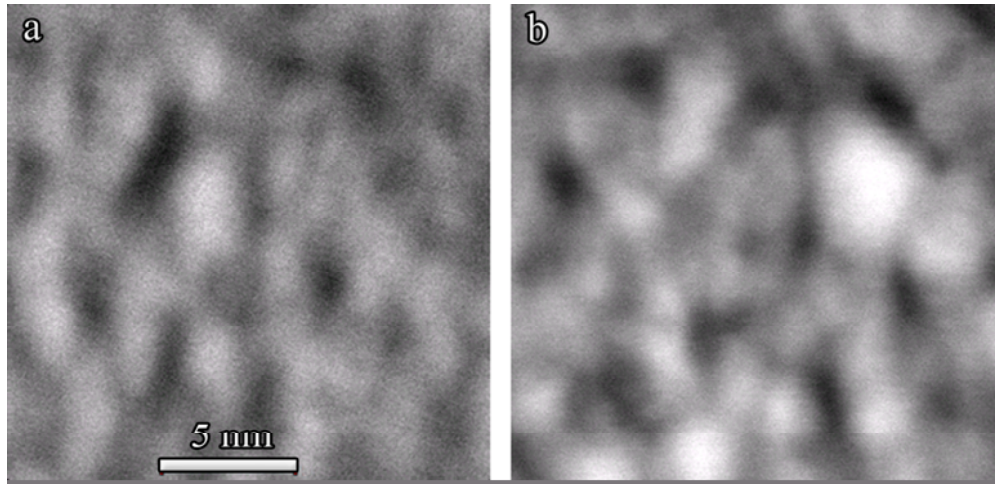

Fig.1

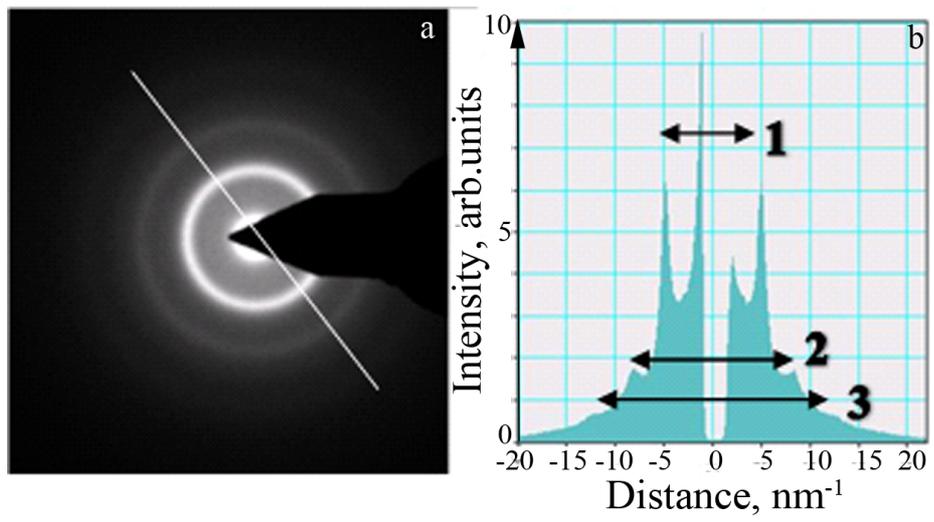

Fig.2

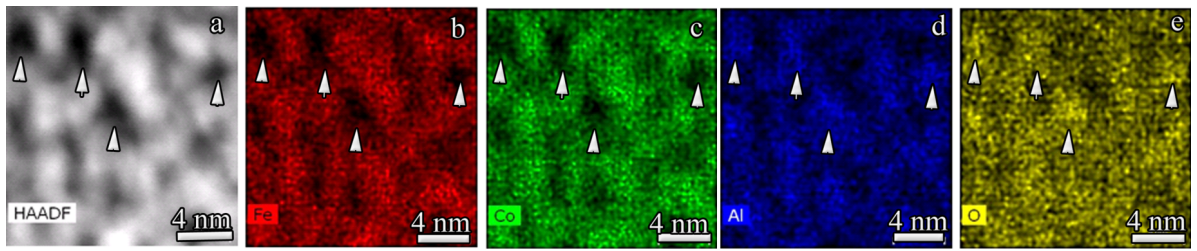

Fig.3

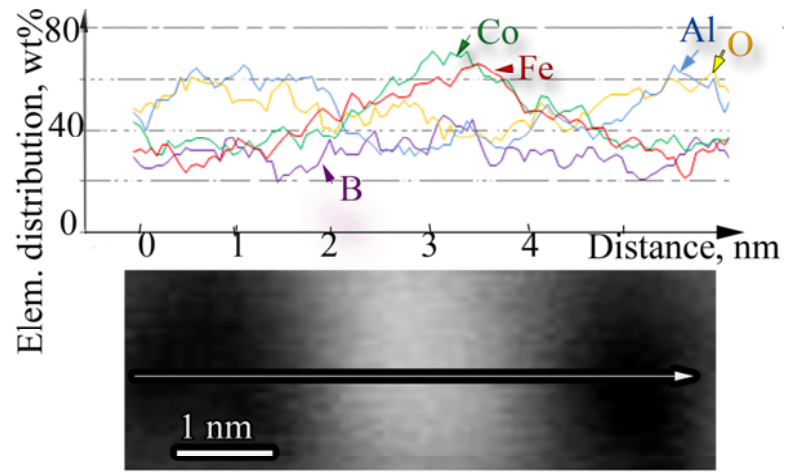

Fig.4.

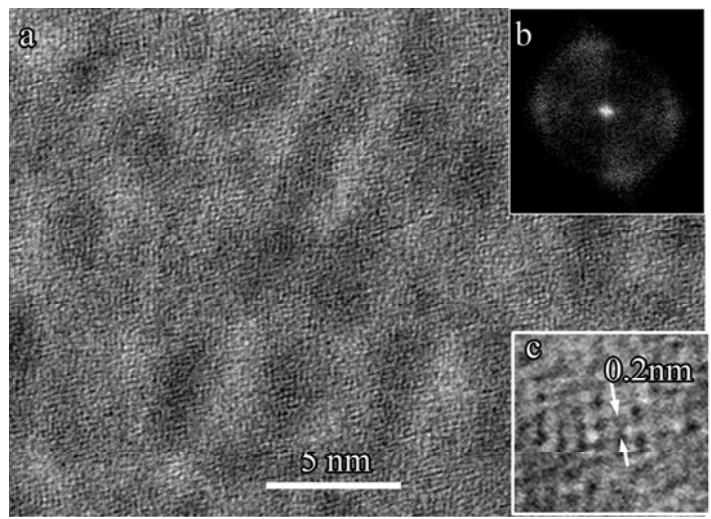

Fig.5

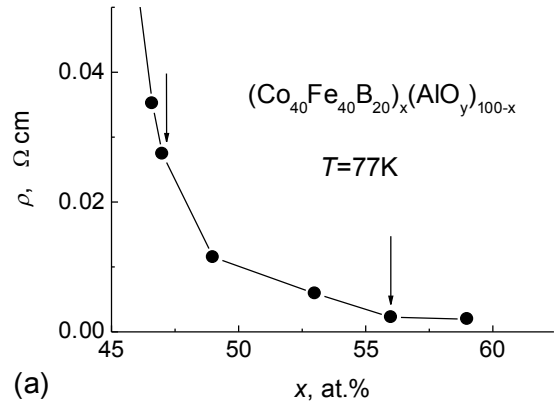

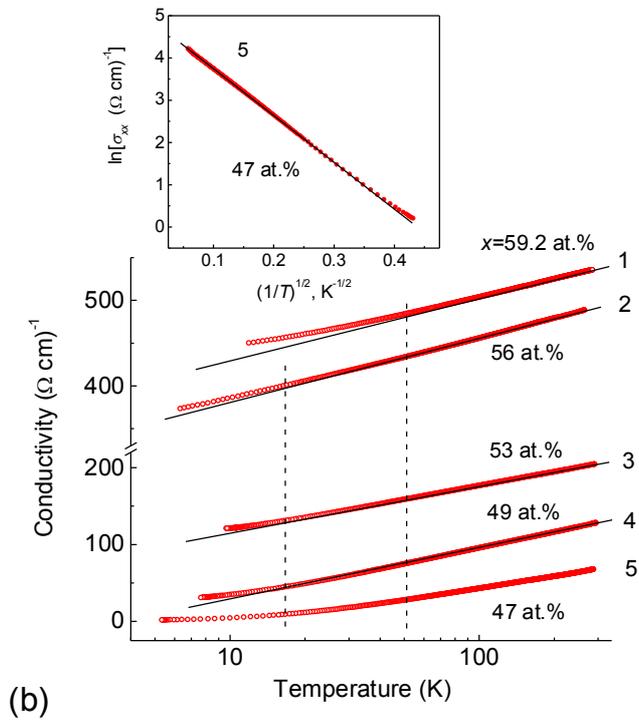

Fig.6

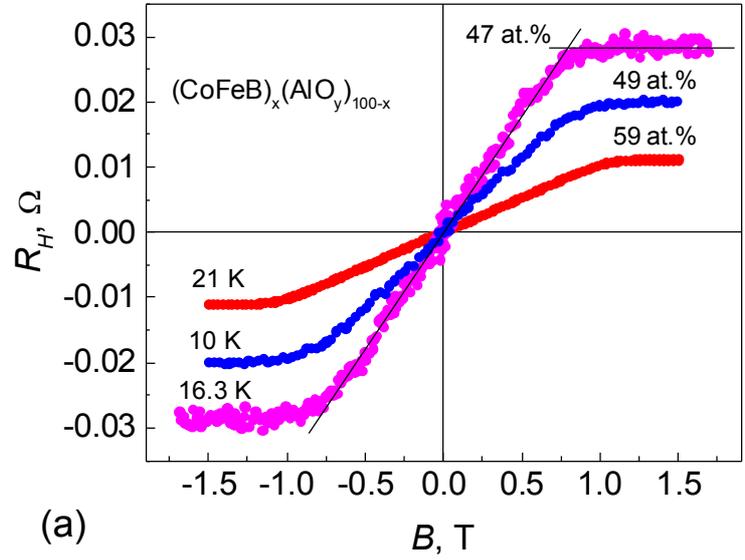

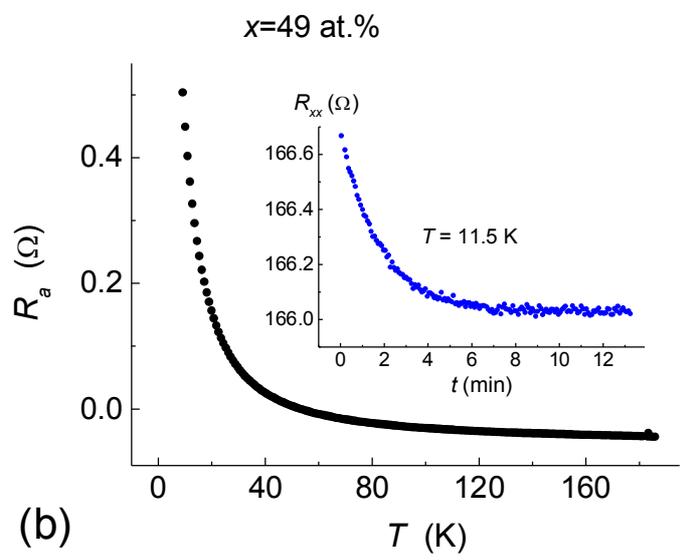

Fig.7

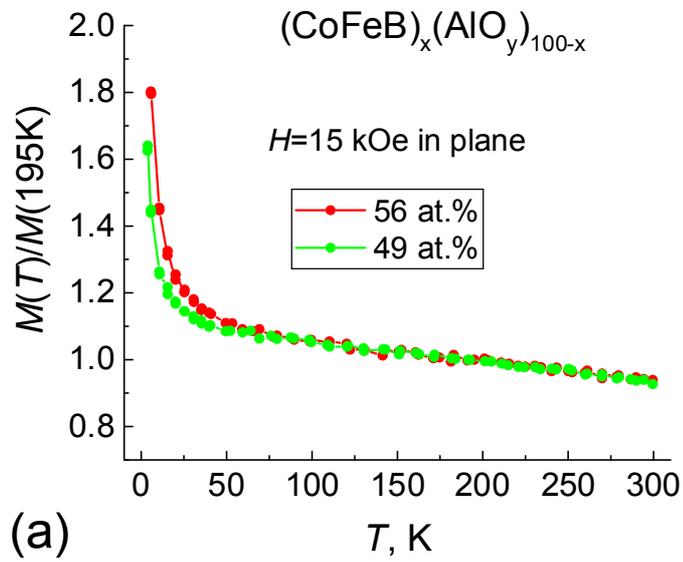

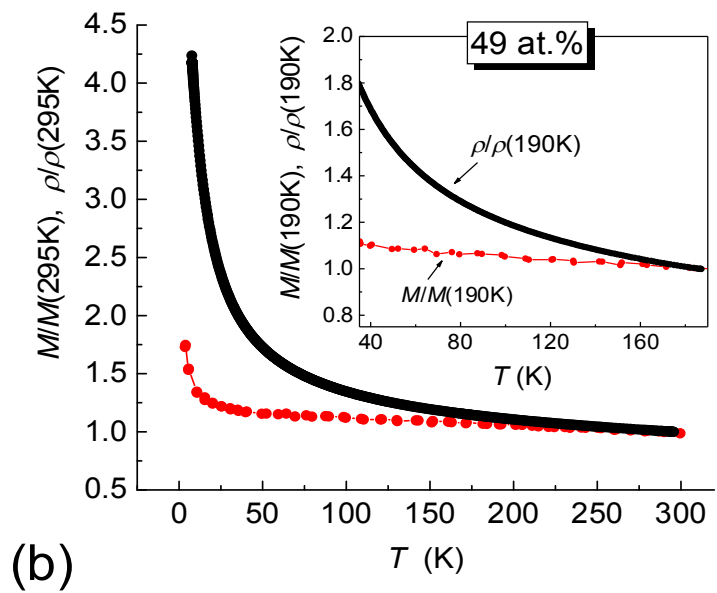

Fig. 8

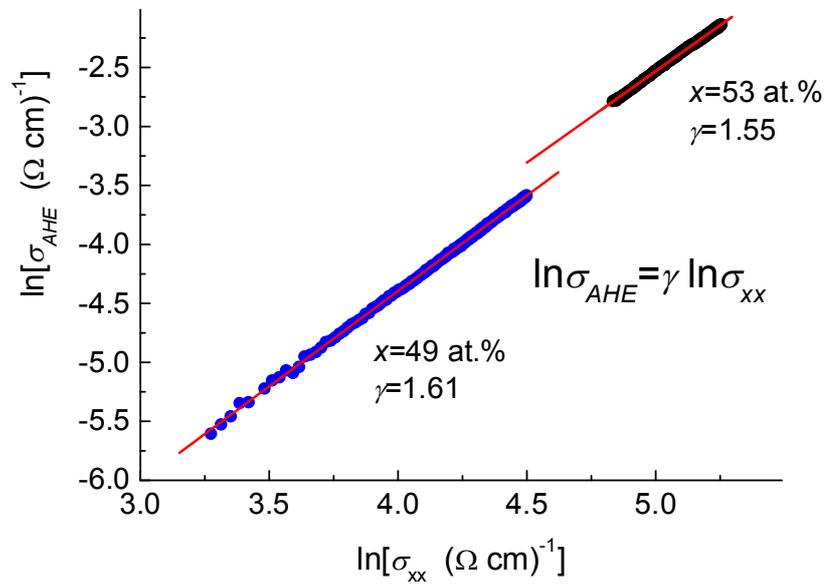

Fig.9

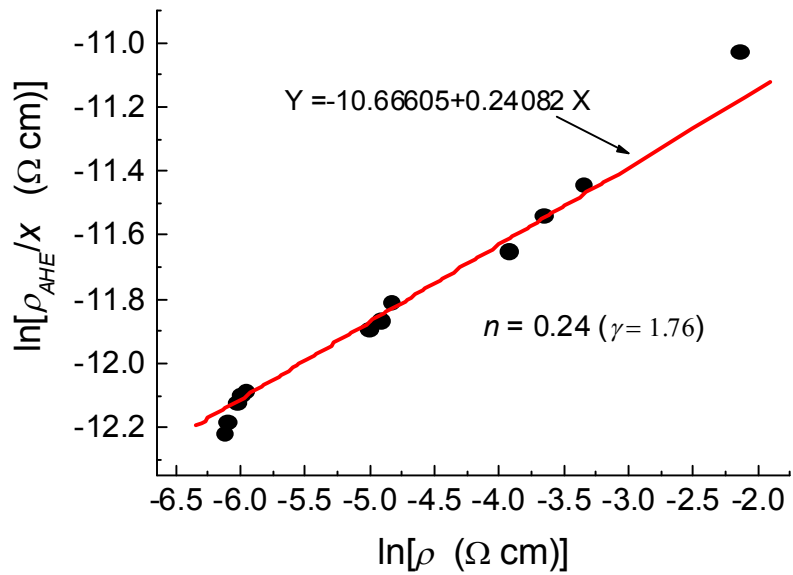

Fig.10

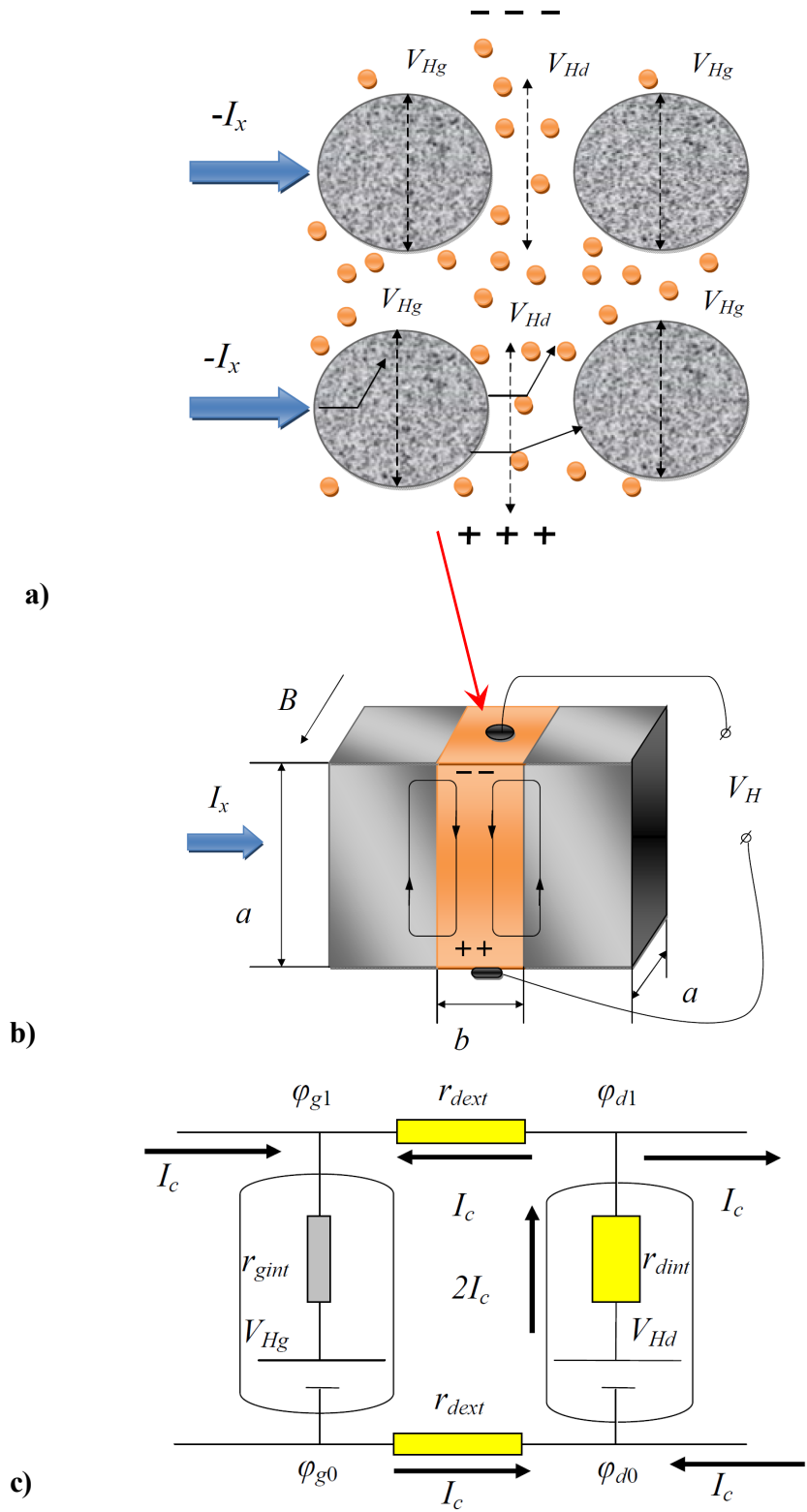

a)

b)

c)

Fig.11

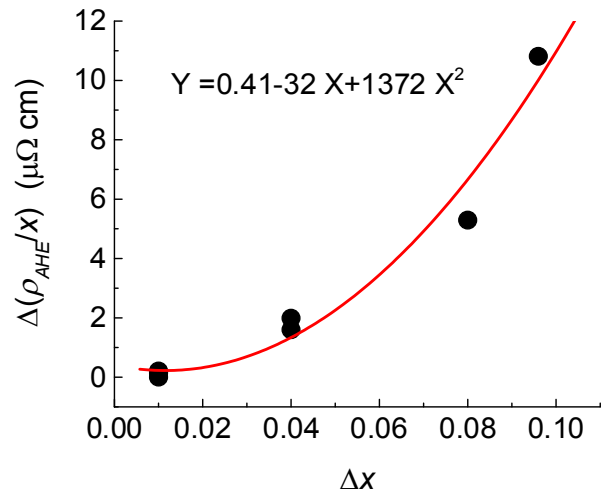

Fig.12